\documentclass[12pt]{article}
\usepackage{amsmath}
\pdfoutput=1
\usepackage{graphicx,psfrag,epsf}
\usepackage{enumerate}
\usepackage{natbib}
\usepackage{url} 
\usepackage{footnote}
\usepackage{url} 
\usepackage{soul}
\usepackage{hyperref}
\usepackage{blindtext}
\hypersetup{colorlinks=true, linkcolor=blue, citecolor=blue, filecolor=blue}
\usepackage{multirow}
\newcommand{\indep}{\perp \!\!\! \perp}
\usepackage{booktabs}
\usepackage{threeparttable}
\usepackage{tabularx}
\usepackage{subfig}
\usepackage{graphicx,psfrag,epsf, float}

\usepackage{setspace}
\def\bfx{\mathbf x}

\def\bfA{\mathbf A}

\def\bfI{\mathbf I}
\def\bfX{\mathbf X}

\def\bfZ{\mathbf Z}

\def\bfM{\mathbf M}

\def\bfG{\mathbf G}
\def\bfalpha{\boldsymbol \alpha}
\def\bfbeta{\boldsymbol \beta}

\def\bfgamma{\boldsymbol \gamma}

\def\bftau{\boldsymbol \tau}
\def\bfPhi{\boldsymbol \Phi}

\def\bfpi{\boldsymbol\pi}

\def\bfzero{\boldsymbol 0}
\def\bfone{\boldsymbol 1}

\def\cG{\mathcal G}

\begin{document}

\def\spacingset#1{\renewcommand{\baselinestretch}%
{#1}\small\normalsize} \spacingset{1}

\date{}
  \title{\bf Bayesian network mediation analysis with application to brain functional connectome}
  
  	\author{
  	{Yize Zhao$^1$}\footnote{Correspondence should be directed to: Yize Zhao (yize.zhao@yale.edu), 300 George Street, New Haven, CT 06511. },
  	{Tianqi Chen$^1$},
    {Jiachen Cai$^1$},
    {Sarah Lichenstein$^2$},\\
    {Marc Potenza$^2$},
    {and Sarah Yip$^2$}\bigskip\\
    $^1$Department of Biostatistics, Yale University, New Haven, CT\\
$^2$Department of Psychiatry, Yale University, New Haven, CT\\
   }
  \maketitle
  \begin{sloppypar}
\spacingset{1.8}
\bigskip
\begin{abstract}
Brain functional connectome, the collection of interconnected neural circuits along functional networks, is one of the most cutting edge neuroimaging traits, and has a potential to play a mediating role within the effect pathway between an exposure and an outcome. While existing mediation analytic approaches are capable of providing insight into complex processes, they mainly focus on a univariate mediator or mediator vector, without considering network-variate mediators. To fill the methodological gap and accomplish this exciting and urgent application, in the paper, we propose an integrative mediation analysis under a Bayesian paradigm with networks entailing the mediation effect. To parameterize the network measurements,  we introduce individually  specified  stochastic  block  models with unknown block allocation, and naturally bridge effect elements through the latent network mediators induced by the connectivity  weights across network modules. To enable the identification of truly active mediating components, we simultaneously impose  a  feature  selection across network mediators. We show the superiority of our model in estimating different effect components and selecting active mediating network structures. As a practical illustration of this approach's application to network neuroscience, we characterize the relationship between a therapeutic intervention and opioid abstinence as mediated by brain functional sub-networks.
\end{abstract}

\noindent%
\vfill

\newpage

\section{Introduction}\label{sec1}
In the past few decades, the investigation of brain functional organization through a collective set of functional brain connections, known as the functional connectome or connectivity, is a rapidly growing research area that has provided novel insights into large-scale neuronal communication and how individual differences in brain functional networks relate to human behavior and psychiatric disorders. With the latest development of modern brain imaging technology, functional imaging such as positron emission tomography (PET) and functional magnetic resonance imaging (fMRI) starts to propagate in behavior and other clinical studies. These provide unique opportunities for integrating complex, whole brain network measures with complex effect pathway analyses, such as those uncovering the underlying effect mechanism between an exposure and an outcome.

This work is motivated by a recent study on opioid use disorder (OUD) \citep{carroll2017galantamine} to investigate the impact of a designed treatment on the opioid abstinence. Within the study, 74 OUD patients were recruited from a randomized controlled trial to receive either behavioral therapy plus galantamine or placebo treatment. At the end of the trial, fMRI data were acquired across the whole brain over multiple scans during both resting state and reward (Monetary Incentive Delay) tasks, and were subsequently transferred into brain functional connectomes  for each participant. Our goal here is to quantify the effect mechanism in the context of mediation framework by explaining how the therapy impacts the post-treatment abstinence mediated by the interconnected brain circuits along functional networks.

Mediation analysis was initially introduced in social and behavioral research and gradually became a popular analytical tool in other disciplines, including the causal inference area in statistics \citep{pearl2000models,vanderweele2011causal}. Under certain assumption regulations, a traditional mediation model investigates two effect paths--a direct one straightly from the exposure (e.g., treatment) to the outcome (e.g., abstinence), and an indirect one intervened by an additional variable, known as a mediator. Among existing mediation analyses, most of them focus on a single mediator variable and characterize different effects through classical linear regressions \citep{baron1986moderator}. Lately, with the emergence of more complex data structure, mediation analysis has been extended to handle multivariate \citep{wang_2013,imai2013identification,daniel2015causal, taguri_2018,vansteelandt_2017,kim2019bayesian} and high dimensional mediators \citep{huang_and_pan_2016,derkach2019high,song2020bayesian} to tackle specific applications. Particularly, in the field of brain imaging, there is a growing trend  to use quantitative neuroimaging measurements as mediators to study the effect mechanism in order to reveal the neuropsychological process after an intervention and how it further modifies human behavior, thus informing future clinical strategies. For example, \cite{lindquist2012functional} considered regional fMRI time series as an imaging mediator and extended mediation model to functional data; \cite{zhao2019granger} further extended the applicable capacity to multiple time series courses. Given the large-scale nature of imaging data, attempts have also been made to handle mediation analysis with high-dimensional mediator vector \citep{chen2018high}.  Within all these studies, the considered mediators are concentrated on brain regional measurements which may be correlated but essentially act as distinct units to characterize intermediated variation. However, given the intricate morphology of human brain, particularly elucidated by the neuronal processing interconnectivity, investigating the mediation role neural system plays based on the more state-of-the-art brain connectome traits becomes an exciting and urgent direction.

To this end, we propose here an integrative mediation analysis with networks entailing the mediation effect from the exposure to the outcome. To parameterize network measurements, i.e. functional connectomes in our case, within our unified Bayesian framework, we construct individually specified stochastic block models (SBM) with the unknown connectivity weights across the connectome modules serving as the latent network mediators. To further remove noninformative components and identify reliable mediating sub-networks over whole brain connectomes, we simultaneously impose a variable selection procedure within our Bayesian model to facilitate identification of distinctions between active and inactive mediating effects. It is worth noting that with a growing interest in studying brain connectivity, plenty of works have been conducted to study partial or whole brain connectivity  to predict behaviors or to integrate with molecular data. Among those, a considerable amount of the analyses simplified the network representation into individual connections, leading to a big loss of graphic topological information. In contrast, by retaining the matrix structure of the brain connectome, depending on its role, a number of scalar-to-network and network-to-scalar models have recently been proposed under decomposition \citep{kong2020l2rm}, penalization \citep{relion2019network} or graphic models \citep{xie2020identifying}. Despite those studies considered to link brain connectomes with other data constructs, their objectives and modeling schemes are not applicable to our problem  with connectomes facilitating mediation.

The major contributions of this paper include the following aspects. First, we make the very first attempt to build a mediation analysis with network-variate mediators. Motivated by investigating the mediating effect of brain functional connectomes, we propose a unified Bayesian mediation framework that is generally applicable to study mediated effects within a network format.    
Second, we simultaneously learn the brain intrinsic functional organization in light of their mediating effect. Utilizing weighted SBMs with unknown block structure embedded within the mediation paradigm, the estimated brain network modules identify neural network components that are likely more clinically relevant than those identified using an unsupervised approach.
Finally, we also accomplish selection among network induced mediators, which removes inactive network features and enhances the analytical power. 

The rest of the article is organized as follows. We present our assumptions and model formulation for the network-variate mediators and different effects in  Section \ref{sec: method}, followed by the corresponding Bayesian framework on prior specification and posterior inference in Section \ref{sec:bayesian}. We assess the model performance by simulations in Section \ref{sec:simulation}, and apply the method to the OUD study in Section \ref{sec:application}. We conclude the paper with a discussion in Section \ref{sec:dis}.

\section{Methods}\label{sec: method}
We start with data structure. For subject $i, i=1,\dots,N$, let $y_i$ denote the outcome, $X_i$ a set of $P$ clinical covariates, and $z_i$ the binary treatment.  For instance, in the OUD study, $y_i$ summarizes the urine test results, $X_i$ includes age and gender, and $z_i$ indicates receiving the treatment or placebo. For each subject, task-based functional MRI (fMRI) data are acquired $K_i$ times. Then, each fMRI time-series course is summarized into a brain functional connectivity network $\cG_{ik}=(\mathcal{V},\mathcal{E}_{ik})$ with vertex set $\mathcal{V}=\{1,\dots,V\}$ including $V$ brain nodes or regions of interest (ROI) defined by a brain atlas, and edge set $\mathcal{E}_{ik}$ collecting functional connections among the nodes. To facilitate analysis, we further represent network $\cG_{ik}$ by its corresponding weighted adjacency matrix $\bfA_{ik}=(a_{ik,jl})$, which is symmetric with $a_{ik,jl}$ indicating the connectivity strength between nodes $j$ and $l$, $0\leq j,l\leq V$. Of note, one could also dichotomize each element $a_{ik,jl}$ by a certain threshold to indicate the existence of a connection. In our model illustration and numerical studies, we choose the former way to keep the continuous scale, but point out that in the case when brain networks are binary, our method can be adjusted by modifying prior specifications and inference algorithm.

In the above context, our goal is to establish effect pathways from the treatment to the outcome mediated by the repeatedly measured brain networks. However, properly modeling network measurements as mediators within a mediation analysis is a challenging task and has not yet been investigated. Despite one could potentially extract all the unique functional connections over brain and implement existing single- or  multivariate-mediator model frameworks, such operations will cause an obvious information loss by completely overlooking the topological configuration within brain networks. In addition, given functional connectivity is non-sparse, the number of unique connections to be considered is $V(V-1)/2$. In brain imaging applications, most of the brain atlases contain at least moderate number of nodes $V$ (e.g. $V=268$ in our application). Directly modeling such a large number of edges as mediators will unavoidably bring prohibitive computation complexity along with hurdles in multiple comparison or model specification. Motivated by those analytical challenges, we propose here an integrative Bayesian mediation analysis to characterize the causal effects among the outcome, treatment and network mediators where the networks are hierarchically modeled by joint SBMs to provide latent mediator surrogates. We will show that our modeling framework admits a desired format to quantify individual path of effect and induce a sub-network level selection on mediation effects.

\subsection{Latent mediators induced by the SBMs}
Converging evidence reveals that brain functional organization encompasses the cognitive processes through sub-networks or sub-components \citep{wig2017segregated}. This supports our anticipation that brain functional connectome is engaged in the mediation pathways through network modules. Along previous literature, though a number of brain sub-network parcellations under resting or task-based functional connectivity have been constructed \citep{yeo2011organization,hamdi2019identification}, they are formed without the supervision of specific clinical procedure. To enhance analytical power and establish more tractable mediation effects, without pre-specifying the sub-community structure, we   assume  $\mathcal{V}$ can be divided into $Q$ unknown latent blocks.  For each node $v\in\mathcal{V}$, we introduce a random vector $G_v=(g_{v1},\dots,g_{vQ})$  to capture community allocation with latent indicator $g_{vq}=1$ if node $v$ belongs to block $q$. Thus, $G_v$ follows a multinomial distribution $G_v\sim \mbox{Multi}(1, \bfpi)$ with $\bfpi=(\pi_1,\dots,\pi_Q)$ the vector of allocation probabilities. 

With $K_i$ functional connectomes collected for subject $i$, for each of the connectivity matrices $\bfA_{ik}$, we construct a weighted SBM with individual edge followed a Normal distribution given the community structure
\begin{equation}\label{eq: SBM}
   a_{ik,jl} \mid g_{jq}=1, g_{lr}=1 \sim \mbox{N}(m_{ik,qr}, \sigma^2_{qr}),
\end{equation}
for $i=1, \dots, N; k=1,\dots, K_i; 1 \le j\neq l \le V; 1 \le q, r \le Q$. Here, $m_{ik,qr}$ is the expected subject/measurement-level  connection strength between blocks $q$ and $r$; and $\sigma^2_{qr}$ is the variance. Due to the symmetric structure, we have $m_{ik,qr}=m_{ik,rq}$ and $\sigma^2_{qr}=\sigma^2_{rq}$. Based on model \eqref{eq: SBM}, connections within the network become independent given the block membership of the nodes they link with, and essentially, it allows us to transfer the representation of each connectivity matrix to a number of sub-network modules with the topological organization absorbed in the community matrix represented by  $\bfG=(G_1,\dots,G_V)$, and the modular characteristics quantified by the strength parameter $m_{ik,qr}$ and variance $\sigma^2_{qr}$ for each block.  It is worth noting that model \eqref{eq: SBM} introduces sub-network encoding for each individual fMRI course within each subject, which is different from most of the SBM related works that only focus on the overall network or multi-subject setting but single modular parameters. A recent paper by \cite{zhang2019mixed} is the only attempt to the best of our knowledge that formulated subject-specific SBMs. However, their work targets on estimating populational connectivity change over time, which is completely different from the problem we are going to address.

Similar to the block-wise connection probability in the traditional SBM, $m_{ik,qr}$ plays an essential role to carry out connectivity variation within and between subjects. When we further assume $m_{i,qr}$ as the subject-level connection strength between regions $q$ and $r$, we have 
\begin{equation}\label{eq: level}
   m_{ik,qr} \sim \mbox{N}(m_{i,qr}, \omega^2_{qr}),
\end{equation}
with $\omega^2_{qr}$ capturing the within subject variance. Conditional on $\bfG$, for each subject, all the unique $m_{i,qr}, 1 \le q \le r \le Q$ can be independently laid out to characterize the modular-wise connection strength. 
As a consequence, parameterizations \eqref{eq: SBM} and \eqref{eq: level} pave a way to facilitate the characterization of network-variate mediators via these unobserved sub-network connectivity, which we call the latent network mediators. We borrow the word ``latent'' from \cite{albert2016causal} to distinguish our proposed mediating variables $M_i=(m_{i,qr}, 1 \le q \le r \le Q)$ with the commonly used fully observed mediators. With the regulation from brain connectivity topology completely undertaken by the modular allocation, the latent network mediators within $M_i$ are conditionally independent. As a valuable property, such independence stabilizes our model fitting \citep{chen2018high} for the conditional expectation of individual mediator.

\subsection{Mediation model}
\begin{figure}
\begin{minipage}{0.43\textwidth}
 \includegraphics[width=2.6in]{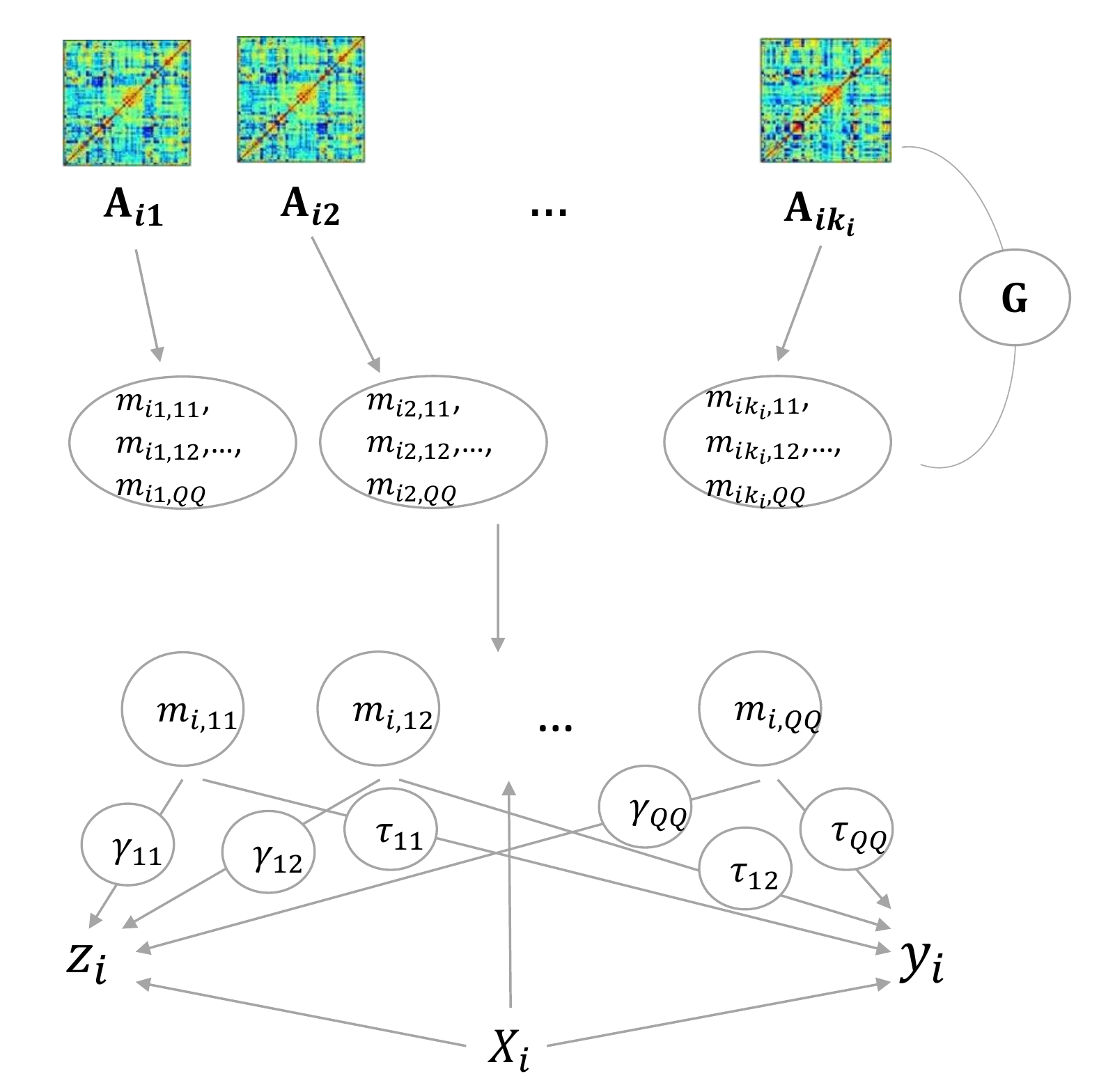} 
\end{minipage}
\begin{minipage}{.55\textwidth}
\begin{align*}
  a_{ik,jl} & \mid g_{jq}=1, g_{lr}=1 \sim \mbox{N}(m_{ik,qr}, \sigma^2_{qr}); \\
 m_{ik,qr} & \sim \mbox{N}(m_{i,qr}, \omega^2_{qr});\\
 G_v & \sim \mbox{Multi}(1, \bfpi);\\
 \bfpi & \sim \mbox{Dir}(p_1,\dots,p_Q)\\
 \gamma_{qr} & \sim \mbox{Bernoulli}(p_{\gamma});\\
 \tau_{qr} & \sim \mbox{Bernoulli}(p_{\tau});\\
 \alpha_{z,qr} &\sim \mbox{N}(0,\sigma^2_{zm});\\
 \beta_{m,qr}& \sim \mbox{N}(0,\sigma^2_{my});\\
 \bfbeta_x& \sim \mbox{N}(\bfzero, \sigma^2_{xy}\mbox{I}) \\
 \beta_z& \sim \mbox{N}(0, \sigma^2_{zy})\\
 \bfalpha_{x,qr}&\sim \mbox{N}(\bfzero, \sigma^2_{xm}\mbox{I});\\
 \sigma^2_{qr}&\sim \mbox{IG}(a_1,b_1);\\
 \omega^2_{qr}&\sim \mbox{IG}(a_2,b_2);\\
 \sigma^2_1, \sigma^2_2, \sigma^2_{zm},  \sigma^2_{my} &\sim \mbox{IG}(0.1, 0.1)\\
   i&=1, \dots, N; k=1,\dots, K_i;\\& 1 \le j\neq l \le V; \\
   & 1 \le q, r \le Q; 1 \le v \le V\\
\end{align*}
\end{minipage}
\caption{A demonstration of the data and parameter structure including repeated measured brain networks, latent mediators induced by SBMs, and the core BNMM modeling framework to identify and characterize the effect components induced by the identified brain subnetworks under each effect pathway.} \label{fig:demo}
\end{figure}

Under the assumption that latent mediatior set $M_i$ directs partial treatment effect, the relationship among the treatment $z_i$, outcome $y_i$, covariates $X_i$, brain connectivity matrices $\{\bfA_{ik}\}_{k=1}^{K_i}$, and their induced latent mediatiors $M_i$ is displayed  inside of Figure \ref{fig:demo}. To formally establish the causal effects, under a counterfactual representation, we denote $M_i(Z)$ the vector of latent mediator values when treatment $z_i=Z$, and $Y_i(Z, M^*_i)$ is the outcome value when $z_i=Z$ and latent mediators $M_i=M^*_i$. Based on them, the natural direct effect (NDE), natural indirect effect (NIE) and total effect (TE) can be defined by
\begin{align}\label{eq: effect}
\begin{split}
\mbox{NDE}&=E(y_i(Z, M_i(Z^*)))-E(y_i(Z^*, M_i(Z^*))),\\ 
\mbox{NIE}&=E(y_i(Z, M_i(Z)))-E(y_i(Z, M_i(Z^*))), \\
\mbox{TE}&=E(y_i(Z, M_i(Z)))-E(y_i(Z^*, M_i(Z^*))),    
\end{split}
\end{align}
with NDE measuring the expected change on outcome by switching treatment from $Z^*$ (e.g. control) to $Z$ (e.g. drug) while maintaining mediators as the original values; NIE quantifying the expected change on outcome when mediators change from $M_i(Z^*)$ to $M_i(Z)$ while fixing treatment; and TE characterizing the overall change on outcome by switching treatment with $\mbox{TE}=\mbox{NDE}+\mbox{NIE}$.
In order to identify NDE and NIE from \eqref{eq: effect}, a number of assumptions known as the sequential ignorability assumptions \citep{imai2010general} are required for the latent mediators. Specifically, we assume that (1) $y_i(Z,M_i)\indep z_i \mid X_i$, (2) $M_i(Z)\indep z_i \mid X_i$,  (3) $y_i(Z,M_i)\indep M_i(Z^*) \mid X_i$, (4) $y_i(Z,M_i)\indep M_i(Z)\mid z_i, X_i$. These assumptions describe that no unmeasured confounding for the relationships between outcome and treatment, mediators and treatment, and outcome and mediators with and without additional controlling of treatment. This set of assumptions serves as the standard requirements for a mediation paradigm and has already been extensively discussed previously including those involving multivariate and high dimensional mediators \citep{wang_2013,song2020bayesian}, latent mediators \citep{derkach2019high}, and neuroimaging applications \citep{lindquist2012functional,chen2018high}. Notably, as discussed in Chen et al. \citep{chen2018high}, though it is extremely hard to rigorously validate the sequential ignorability assumptions in the real practice, when one or more of the assumptions fail to hold, we can still use \eqref{eq: effect} to quantify potential mediating effects in exploratory analysis. In addition, for our application, though connections within a brain connectome are correlated with each other, our constructed latent mediators are conditionally independent based on the SBM assumption. Therefore, despite our main objective is to uncover the average effect pathway bridging the treatment, brain activation along connectomes, and outcome, we do have the capacity to characterize the modular specific indirect effects.

Given both our outcome and latent mediators are Normally distributed, we build regression models for the conditional distributions $y_i\mid M_i, X_i, z_i$ and $m_{i,qr}\mid X_i, z_i, 1 \le q \le r \le Q$. Previous empirical studies on individual connections indicate that a certain proportion of connectivity features within brain indeed admit negligible variations between the treatment and placebo group \citep{lichenstein2019dissociable}. Some recent works on connectome-based prediction also reveal that several key sub-networks play a dominate role in predicting people's behavior \citep{shen2017using}. The complication of brain neural circuits with respect to their impact on behavior and correspondence to exposure motivates us to make the following biologically plausible modeling assumptions: first, only a subset of brain network modules and their induced latent network mediators are significantly impacted by treatment; second, only a subset of brain network modules and their induced latent network mediators play a significant role in influencing outcome. To this end, we have the following sparse regressions 
\begin{align}\label{eq: model1}
\begin{split}
y_i&=X^T_i\bfbeta_x+M^T_i(\bfbeta_m\circ\bftau)+z_i\beta_z+\epsilon_{i} \\
m_{i,qr}&=X^T_i\bfalpha_{x,qr}+z_i(\alpha_{z,qr}\cdot\gamma_{qr})+\psi_{i,qr}, \quad 1 \le q \le r \le Q,
\end{split}
\end{align}
where $X_i$ includes 1 as the first element; $(P+1)\times 1$ vector $\bfbeta_x$, $Q(Q+1)/2\times 1$ vector $\bfbeta_m=(\beta_{m,qr},1 \le q \le r \le Q)^T$ and scalar $\beta_z$ represent the effects of covariates, mediators and treatment on the outcome; $(P+1)\times 1$ vector $\bfalpha_{x,qr}$ and scalar $\alpha_{z,qr}$ are the coefficients for covariates and treatment on mediator $m_{i,qr}$; and there are random errors $\epsilon_i\sim \mbox{N}(0,\sigma_1^2)$, $\psi_{i,qr}\sim \mbox{N}(0,\sigma_2^2)$. Inside the models \eqref{eq: model1}, we also include latent selection indicator sets $\bftau=(\tau_{qr}\in\{0, 1\}, 1 \le q \le r \le Q)$ and $\bfgamma=(\gamma_{qr}\in\{0, 1\}, 1 \le q \le r \le Q)$, both with a dimension $Q(Q+1)/2\times 1$ to impose the desired sparsity, where $\circ$ denotes the entry-wise product. Specifically, we have $\tau_{qr}=1$ if latent mediator $m_{i,qr}$  brings significant impact to outcome, 0 otherwise; and $\gamma_{qr}=1$ if treatment  effect is significant on mediator $m_{i,qr}$, 0 otherwise. Given the fact that brain connectivity modules influenced by the treatment and those impacting the outcome can be different, to allow flexibility, we use separate selection indicator sets here without any constrain to match between $\bftau$ and $\bfgamma$.

Based on models \eqref{eq: effect} and \eqref{eq: model1}, under the above sequential ignorability assumptions, the average direct and indirect effects can be identified and expressed as
\begin{align}\label{eq: effect2}
\begin{split}
\mbox{NDE}&=\beta_z(Z-Z^*),\\ 
\mbox{NIE}&=(Z-Z^*)\sum_{q \le r}\sum_{r \le Q}(\alpha_{z,qr}\cdot\gamma_{qr})(\beta_{m,qr}\cdot\tau_{qr}), \\
\mbox{TE}&=(Z-Z^*)\{\beta_z+\sum_{q \le r}\sum_{r \le Q}(\alpha_{z,qr}\cdot\gamma_{qr})(\beta_{m,qr}\cdot\tau_{qr})\}.
\end{split}
\end{align}
According to the above representation, a latent mediator $m_{i,qr}$ will contribute to both NIE and TE only if both $\tau_{qr}$ and $\gamma_{qr}$ are nonzero. In other words, the truly active latent mediators are the ones that are jointly selected in both models in \eqref{eq: model1}. \cite{song2020bayesian} also discussed different patterns of activation within the potential mediators. In their paper, they use shrinkage priors on coefficients to regulate small effects to achieve quasi-sparsity and define active mediators as the ones with both related coefficients coming from the larger Normal components. Here, we directly impose sparsity using selection indicators, which leads to an explicit definition for  each latent network mediator's activation based on its inclusion or exclusion from the averaged effects. 

\section{A unified Bayesian framework}\label{sec:bayesian}
To jointly model stochastic block structure of functional connectome and its associated mediation analysis, we develop a Bayesian estimation and inference framework for models  \eqref{eq: SBM}, \eqref{eq: level} and \eqref{eq: model1}. This enable us to overcome the difficulty in performing inference under a conventional frequentist mediation analysis \citep{yuan2009bayesian}, and effectively integrate the construction of mediated brain module  within the effect pathway establishment. 
\subsection{Prior specification}
We assign prior for each unknown parameter within the proposed hierarchical models. For the SBM related parameters, given  $G_v$ follows a Multinomial distribution with probability $\bfpi$, we have $\bfpi$ follow a conjugate Dirichlet distribution $\bfpi\sim \mbox{Dir}(p_1,\dots,p_Q)$. Each within block variance is set to follow an Inverse Gamma (IG) distribution $\sigma^2_{qr}\sim \mbox{IG}(a_1,b_1)$, and each inner measurement variance $\omega^2_{qr}\sim \mbox{IG}(a_2,b_2)$. To determine the total number of blocks $Q$, a number of tools have been developed for SBMs using different evaluation metrics such as hypothesis testing  \citep{bickel2016hypothesis}, cross validation \citep{chen2018network}, and Bayes Factor  \citep{bayes_factor}, etc. Through our analytical experiments, we determine $Q$ based on the Integrated Completed Likelihood (ICL) criterion  \citep{blockmodels_icl}, which shows a more robust performance in our numerical studies. 
In terms of the parameters involved in the regression models, we assign Bernoulli distributions for each side of the mediating effect selection indicators $\gamma_{qr}$ and $\tau_{qr}$, with probabilities $p_{\gamma}$ and $p_{\tau}$, respectively. For the regression coefficients, we assume each $\alpha_{z,qr} \sim \mbox{N}(0,\sigma^2_{zm}), \beta_{m,qr} \sim \mbox{N}(0,\sigma^2_{my})$, 
and the coefficients $\bfbeta_x\sim \mbox{N}(\bfzero, \sigma^2_{xy}\mbox{I})$, $\beta_z\sim \mbox{N}(0, \sigma^2_{zy})$ and each $\bfalpha_{x,qr}\sim \mbox{N}(\bfzero, \sigma^2_{xm}\mbox{I}), 1 \le q \le r \le Q$. At last, we impose noninformative IG priors with  shape and scale parameters set to be 0.1  for 
$\sigma^2_1, \sigma^2_2, \sigma^2_{zm},  \sigma^2_{my}$; and set
 $\sigma^2_{xy},\sigma^2_{zy},\sigma^2_{xm}$ to be large values e.g. 10. 
We name our unified Bayesian mediation analysis the \textit{Bayesian network mediation model} (BNMM) and the core modeling structure along with all the prior specifications is further included to Figure \ref{fig:demo}.

\subsection{Posterior inference}
Given the outcome, exposure, repeatedly measured brain connectomes and covariates, we first write down the joint conditional posterior distribution for the unknown parameters $\bfPhi=(\bfG,\bfbeta_x,\bfbeta_m,\bftau,\bfgamma, \beta_z,\bfalpha_{x,qr},\bfalpha_{z,qr},\sigma^2_1,\sigma^2_2,M_i,m_{ik,qr},\sigma^2_{qr},\omega^2_{qr},  k=1,\dots,K_i;i=1,\dots, N;1 \le q \le r \le Q)$ given the observed data

\begin{align*}
\begin{split}
&f\{\bfPhi\mid (y_i, z_i, X_i, \bfA_{ik}, k=1,\dots,K_i;i=1,\dots, N)\}\propto \prod_{i=1} f(y_i\mid z_i, X_i, M_i,\bfbeta_x,\bftau,\beta_z,\sigma_1^2)\\ &\prod_{i=1}\prod_{k}  \prod_{1 \le q \le r \le Q} f(\bfA_{ik}\mid m_{ik,qr}, \sigma^2_{qr}, \bfG)f(m_{ik,qr}\mid m_{i,qr},\omega^2_{qr})f(m_{i,qr}\mid X_i,z_i,\bfalpha_{x,qr},\alpha_{z,qr},\gamma_{qr},\sigma^2_2)\\
& \prod_{1 \le q \le r \le Q}\prod_i\prod_{k}f(\sigma^2_{qr}) f(\omega^2_{qr})f(\bfalpha_{x,qr})f(\alpha_{z,qr})f(\gamma_{qr})f(\bfG)f(\bfbeta_x)f(\beta_z)f(\bftau)f(\sigma_1^2)f(\sigma^2_2).
\end{split}
\end{align*}

We rely on a Markov Chain Monte Carlo (MCMC) algorithm to conduct the posterior inference. Since we can directly obtain the analytical form of the full conditional distribution of each unknown parameter, the computational cost is not heavy through Gibbs sampler.  We briefly describe the sampling steps for each MCMC iteration here and provide the full algorithm in the supporting web materials.

\begin{itemize}
  \item Given the current value of $\bftau$, split $\bfbeta_m$ into $\bfbeta_{m0}$ and $\bfbeta_{m1}$ corresponding to the unselected ($\bftau=\bfzero$) and selected ($\bftau=\bfone$) brain network blocks for the outcome regression. Update $\bfbeta_{m0}\sim \mbox{N}(\bfzero,\sigma^2_{my}\bfI)$; and $\bfbeta_{m1}\mid \bfM, \bfbeta_x,\beta_z, \sigma^2_{my},\sigma^2_1$ from its posterior multivariate Normal distribution, where $\bfM=(M_1,\dots,M_n)^T$.
  \item Given the current value of $\{\gamma_{qr}\}$, denote $\bfalpha_{z0}=(\alpha_{z,qr}: \gamma_{qr}=0)$ and $\bfalpha_{z1}=(\alpha_{z,qr}: \gamma_{qr}=1)$ corresponding to brain network blocks receiving or not receiving treatment impact. Update   $\bfalpha_{z0}\sim\mbox{N}(\bfzero,\sigma^2_{zm}\bfI)$ and $\bfalpha_{z1}\mid \bfalpha_x, \bfM, \sigma^2_{zm},\sigma^2_2$ from its posterior multivariate Normal distribution.
  \item Define $l_{\tau}(\tau_{qr}\mid \bfM, \bftau_{-qr},\bfbeta_m,\bfbeta_x,\beta_z,\sigma^2_1):=\prod_{i}\Phi[\frac{y_i-\bfX^T_i\bfbeta_x-M^T_i\{\bfbeta_m\circ(\tau_{qr}~ \bftau_{-qr})\}-z_i\beta_z}{\sigma_1}]$ with $\Phi(\bfx)=\bfx^T\bfx/2$. Update $\tau_{qr}, 1 \le q \le r \le Q$ from a Bernoulli distribution with probability $p_{\tau}l_{\tau}(1)/\{p_{\tau}l_{\tau}(1)+(1-p_{\tau})l_{\tau}(0)\}$. In a similar way, we can define $l_{\gamma}(\gamma_{qr}\mid \bfM, \bfgamma_{-qr},\bfalpha_{x,qr},\bfalpha_{z,qr},\sigma^2_2)$, and update each $\gamma_{qr}\sim \mbox{Bern}(p_{\gamma}l_{\gamma}(1)/\{p_{\gamma}l_{\gamma}(1)+(1-p_{\gamma})l_{\gamma}(0)\})$.
  \item Update nuisance parameters $\bfbeta_x\mid \bfM, \bfbeta_m,\bftau,\beta_z,\sigma^2_1$ and  $\bfalpha_{x,qr}\mid \bfM, \alpha_{z,qr},\bfgamma_{qr},\sigma^2_2$ with $1 \le q \le r \le Q$ from their corresponding posterior multivariate Normal distributions.
  \item Update $\beta_z\mid \bfM, \bfbeta_x,\bfbeta_m,\bftau,\sigma^2_1$ from its posterior Normal distribution.
  \item Update latent mediators $M_i\mid \{m_{ik,qr}\}, \bfbeta_x, \bfbeta_m,\bftau,\beta_z,\{\alpha_{x,qr}\}, \{\gamma_{qr}\},\sigma^2_1,\sigma^2_2,\{\omega_{qr}\}$ with $i=1,\dots,N$ from the posterior multivariate Normal distribution. The conditional subject/measurement-level connection strength $m_{ik,qr}\mid M_i, \bfG, \omega_{qr}, \sigma^2_{qr}$ with $k=1,\dots,K_i, i=1,\dots,N, 1 \le q \le r \le Q$ is updated from the posterior Normal distribution.
  \item Update community allocation $G_v\mid \bfpi, \{m_{ik,qr}\}, \{\sigma^2_{qr}\}$ for $v=1,\dots,V$ from its posterior Multinomial distribution. The hyper-parameter $\bfpi\mid \bfG$ is updated from the posterior Dirichlet distribution.
  \item Update  $\sigma^2_1\mid \bfbeta_x,\bfM, \bfbeta_m,\bftau,\beta_z, Z$; $\sigma^2_2\mid\bfM, \{\bfalpha_{x,qr}\},\bfgamma$; $\omega_{qr}\mid \{M_i\}, \{m_{ik,qr}\}$; $\sigma_{qr}\mid \{\bfA_i\}, \{m_{ik,qr}\}$; $\sigma^2_{zm}\mid\{\bfalpha_{x,qr}\}$ and $\sigma^2_{my}\mid \{\beta_{m,qr}\}$ from their corresponding posterior IG distributions. 
\end{itemize}

For implementation, we start with random initials for multiple chains and check the posterior convergence after completing the inference by both trace plots and GR method \citep{gelman1992inference}. To determine the truly active latent network mediators and the overall  NIE  and  TE,  we take the posterior median model \citep{barbieri2004optimal} corresponding to a 0.5 threshold on the marginal posterior inclusion probabilities of $\bftau$ and $\bfgamma$ for the effect related parameters. The parcellation of brain sub-networks within the mediation pathways are summarized by the posterior mode of each $\bfG_v, v=1,\dots,V$ to map each brain region to its mediating-related module.

\section{Simulation}\label{sec:simulation}

\begin{figure}
    \centering
    \includegraphics[width=6in]{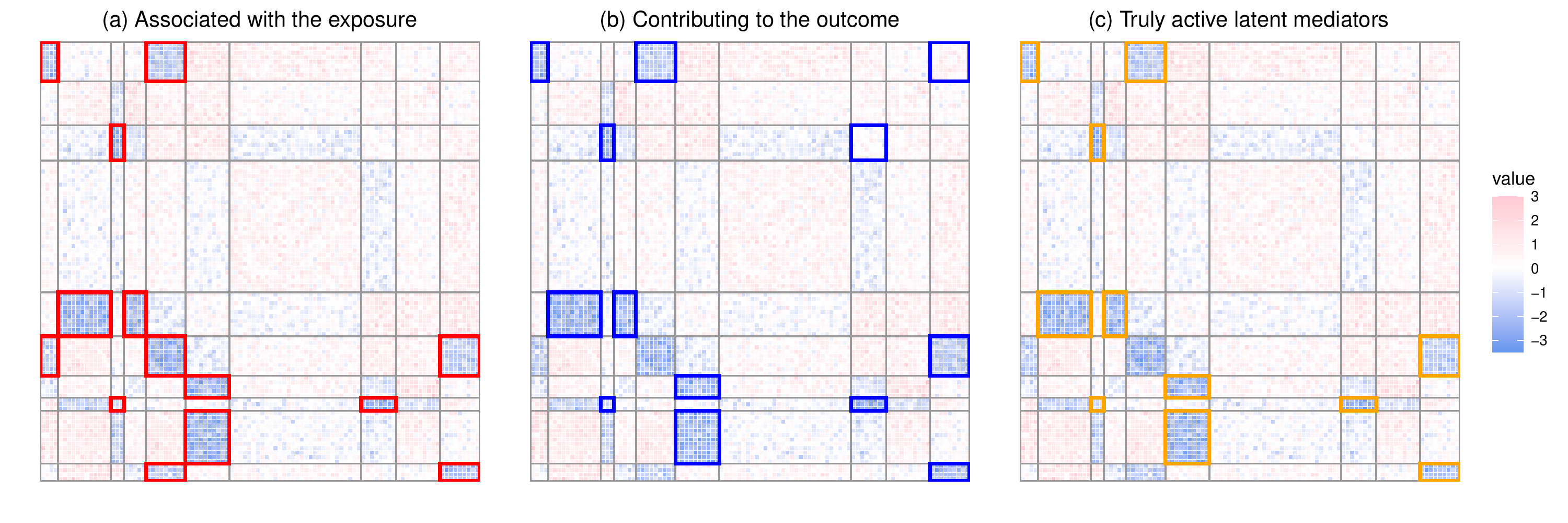}
    \caption{Generated block structure and the informative sub-networks that are highlighted from a randomly selected brain connectome from the Scenario 2. A set of the induced latent network mediators by the informative sub-networks are (a) associated with the exposure, (b) contributing to the outcome, and (c) the truly active mediators. }
    \label{fig:simuset}
\end{figure}

We now evaluate the finite sample performance of our proposed BNMM compared with existing alternatives by simulations. We generate data of 50 subjects, with each subject collected 6 repeated measures on connectome over 100 nodes. We consider both a continuous and a binary exposure, which are primarily distinct on the result interpretation without a significant modification on our model implementation. However, given the implementation for one of the major competing approaches \citep{song2020bayesian} is designed for a continuous exposure, we focus on illustrating this case with the exposure generated from a standard Normal distribution.  To construct latent network mediators, we first set $Q=10$, and generate the community allocation $G_v$ for each node $v$ from a Dirichlet-multinomial distribution with concentration parameters all set to 3. We follow \eqref{eq: model1}, \eqref{eq: level} and \eqref{eq: SBM} to generate the latent network components $\{m_{i,qr}\}$, $\{m_{ik,qr}\}$, outcome $\{y_i\}$ and brain connectivity matrices $\{\bfA_{ik}\}$ where $\beta_{z}=1.5,  \bfbeta_{m}=\mathbf{2}, \bfbeta_{x}=1, \bfalpha_{x,qr}=0.3$,  $\omega_{qr}=0.1$ and $\bfalpha_{z,qr}, 1 \le q \le r \le Q$ range from 1.5 to 2.5. In terms of the signal to noise ratio, we consider a low noise case with $\sigma_1=\sigma_{2}=\sigma_{qr}=0.1$; and a high noise case with $\sigma_2=\sigma_{qr}=0.5, \sigma_{1}=1$. As for the active latent brain sub-networks along the mediation pathways marked by $\bftau$ and $\bfgamma$, we consider two scenarios. In the \textit{first scenario}, we set $\bftau$ and $\bfgamma$ to be the same  which means the latent mediators that impact the outcome are those and only those influenced by the exposure.  This is the prerequisite assumption for any univariate- or multivariate-mediator analysis without imposing mediator selection, but could be unrealistic in clinical studies. In the \textit{second scenario}, we mimic the real world setting where the non-zeros in $\bftau$ and $\bfgamma$ may be overlapped but contain difference. Hence, we allow certain brain sub-networks to soak impact from the exposure without responding to the outcome or intrinsically alter the outcome without interfering by the exposure, and the truly active mediators are the ones hold effects on both directions. Figure \ref{fig:simuset} demonstrates the generated sub-network structure and different signal patterns within a connectome to reflect the above considerations under the second scenario.

We generate 100 Monte Carlo datasets for each setting.  To implement the BNMM, we set noninformative hyper-priors for the variance parameters as mentioned in the posterior computation in addition to $a_1=b_1=a_2=b_2=1$, and assign $p_{\tau}=p_{\gamma}=0.5$. The MCMC is conducted over 5,000 iterations with 2,000 burn-in. In terms of the alternative methods, given none of the existing models can be directly applied to handle network-variate mediators, we ought to transfer connectivity networks into a different format in order to implement competing approaches. Specifically, since our proposed BNMM jointly models the latent mediators derived from observed networks and the effect pathways, for the competing methods, we consider 1) a direct use of the true latent sub-network connectivity strengths as mediators, i.e.  pretending $M_i$ are fully observed with their true values as input; and 2) a direct extraction of unique connections from the averaged brain connectivity matrix over repeated measures as mediators. The mediators are multivariate under both strategies, and we apply an univariate mediation analysis implemented by the R package {\tt Mediation} on each individual mediator sequentially as well as a recent Bayesian mediation model on high-dimensional mediators (BAMA) \citep{song2020bayesian} implemented by the R package {\tt BAMA}. Overall, we implement four additional approaches: univariate  mediation  analysis under true latent mediator (UMLM), BAMA under true latent mediator (BAMALM), univariate  mediation  analysis under brain connections (UMBC), and BAMA under brain connections (BAMABC). Of note, the above strategies weaken the influence from modeling network structure which is infeasible for the competing methods. Particularly, the implementation setting is in favor of the first two competing approaches when the true latent mediators detached from the network configuration are directly imposed, compared with the BNMM with the latent mediators needed to be jointly modeled. 
Finally, we evaluate both the selection of the truly active mediators measured by sensitivity and specificity, and the bias of estimating NDE, NIE and TE. For the selection metrics, to maintain the summary consistent, we always map the selected mediators (sub-network or edges) back to the original connectome domain, and 
all the results are summarized in Table \ref{table:simulation}.

Based on the results, we can conclude the proposed BNMM achieves the best or close to the best performance in identifying truly active mediators and estimating different effect components under all the simulation settings. Specifically, our method achieves a superior accuracy in distinguishing mediators from noises within repeatedly measured brain connectomes as showed by the close to one sensitivity and specificity. When comparing between two scenarios, the nonidentical sub-network effects associated with exposure and outcome in Scenario 2 deteriorate the selection accuracy for all the methods by a more complex effect mechanism. Similarly, the performance expectedly gets worse under high noises. However, among all, BNMM maintains its superior selection performance even under the toughest settings. In terms of the effect estimation, we see from the Table \ref{table:simulation} that BNMM obtains the smallest bias in estimating NDE, NIE and TE under almost all the settings except the Scenario 2 with a large noise, where BNMM  slightly underperforms UMLM in estimating the NIE, and  BAMALM in estimating the NDE. However, UMLM and BAMALM have a much worse performance in other evaluation metrics under this setting. Overall, we demonstrate the superiority and robustness of BNMM under the current simulations. Meanwhile, we also confirm the advantage of joint modeling network mediator and effect pathways given our method generally outperforms UMLM and  BAMALM which directly use the true latent mediators as inputs. In terms of the comparisons among the competing methods, UMLM and BAMALM outperform their alternative implementation under UMBC and  BAMABC, which is anticipated given the former ones directly adopt the true mediator values. Between the univariate mediation analysis and BAMA, BAMA tends to be more conservative with less mediators selected resulting in a smaller sensitivity. The  univariate analysis, though identifying more true positives, suffers with worse performance in effect estimations.

\section{Application for an OUD study}\label{sec:application}

We apply our proposed approach to the motivated OUD study. 
Only participants with acceptable neuroimaging data for both pre- and post-treatment fMRI scans are included here, for a total of 41 subjects (18 active drug and 23 placebo). Preprocessing was conducted as previously described \citep{lichenstein2019dissociable} and included: discarding the first 6 volumes or each fMRI run, skull stripping, slice-time correction, motion correction, temporal smoothing (Gaussian filter with approximate cutoff frequency=0.12Hz), normalization and concatenation of functional task runs, and nonlinear registration to structural data. The following covariates were regressed out of the data: linear and quadratic drifts, mean cerebral-spinal-fluid, mean white-matter signal, and a 24-parameter motion model including six rigid-body motion parameters, six temporal derivatives, and these terms squared \citep{satterthwaite2013improved}. SPM8 (http://www.fil.ion.ucl.ac.uk/spm/) was used for slicetime and motion correction. Additional preprocessing was conducted using BioImage Suite \citep{joshi2011unified}.

For all the subjects, biweekly urine testing was conducted during the trial period and the opioid abstinence outcome was measured by the transferred negative urine specimens test results for non-methadone opioids.  We focus on the reward task for the connectome mediator where each subject has multiple scans  collected  and the number of scans are different among subjects ranging from four to six. To construct functional connectivity, we adopt the Shen 268-node brain atlas that includes cortex, subcortex, and cerebellum \citep{shen2013groupwise}. The task connectivity is defined on the basis of node-by-node pairwise Pearson’s correlations using the raw task time courses among all the brain regions. These $r$ statistics are further transformed to be Normally distributed using Fisher’s $z$-transformation, and for each participant, we summarize each functional connectome collection into a $268\times268$ connectivity matrix. Finally, we also consider age and gender in both the outcome and the latent mediator models as covariates.

\begin{figure}
    \centering
    \includegraphics[width=6in]{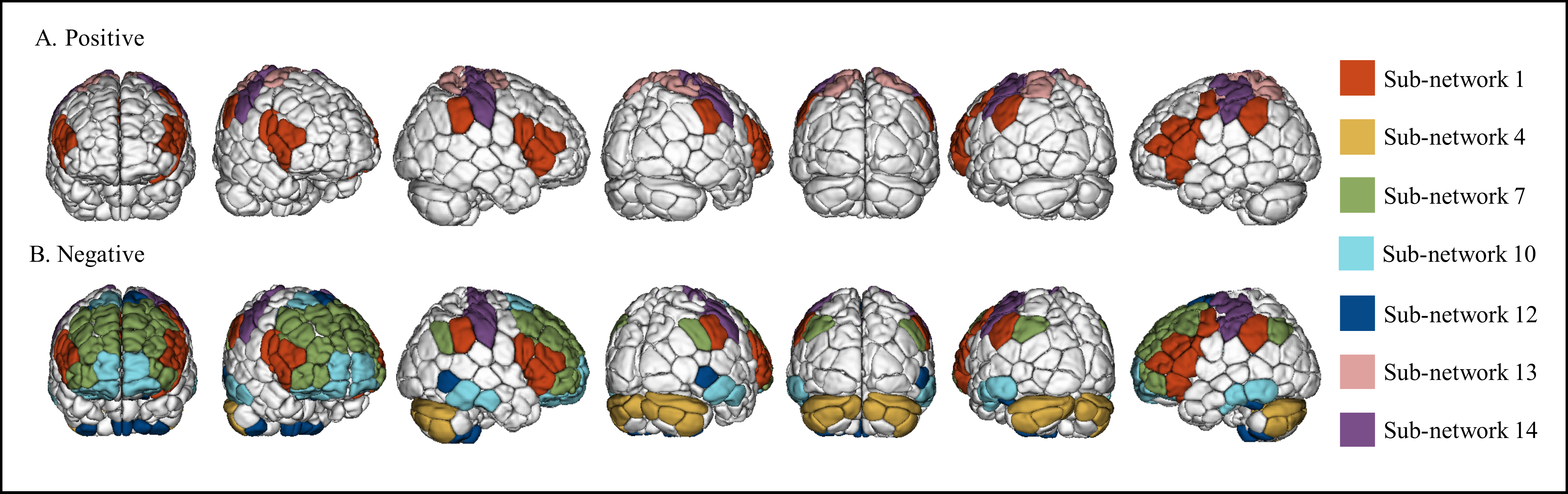}
    \caption{The identified sub-networks within the truly active latent network mediators involving in the positive and negative contributions to the NIE.}
    \label{fig:sub}
\end{figure}

We apply BNMM to jointly dissect the connectivity module topology to assist in mediation analyses and to uncover the effect pathways, in which functional networks mediated relationships between treatment and opioid abstinence. As discussed previously, we determine the block number using the ICL criteria and set $Q=14$. The rest of parameter settings and implementation closely follow the procedure in the simulations. Eventually, based on the posterior samples, the estimated NDE, NIE and TE with 95\% credible interval are 0.59 (0.12, 1.12),  -0.12 (-0.43, 0.66) and 0.48 (0.01, 1.03); and the NIE consists of -0.62 (-1.05, -0.16) for the negative component and 0.51 (0.03, 0.98) for the positive one, aligning with the negative and positive sub-networks typically seen in previous functional connectome analyses. 
We further investigate how the NIE develops through the brain sub-networks. Based on the estimated allocation matrix $\bfG$, the 268 nodes are split into 14 sub-networks within the mediation framework. These sub-networks are further compared with the canonical neural networks \citep{power2011functional}, and in Table \ref{table:real1}, we list the number of nodes within each sub-network, the most overlapping canonical networks, and the number of shared nodes ($N_{c}$), with the complete results provided in the supporting web materials. As can be seen in the table, our constructed sub-networks in the mediation analysis present a complex pattern of sub-networks that are composed of multiple components of canonical neural networks. This is anticipated given that the canonical networks were constructed in an unsupervised fashion using the resting-state functional connectivity for the default cognitive functions, and our organizations are established specifically in relation to a treatment effect mechanism. Among the latent mediator sets formed by the connectivity weights between and within those sub-networks, we eventually identify six active mediators. Specifically, the latent connection strength between sub-network 13 and sub-networks 1 and 14 positively contribute to the NIE; and  the latent connection strength within sub-network 12, between sub-network 4 and sub-networks 7 and 10, and between sub-network 1 and sub-network 14 are negative contributors. We show in Figure \ref{fig:sub} the locations of these sub-networks implicate in the positive and negative components of the NIE. To further investigate their network anatomy, we summarize the included macroscale brain regions  for all the sub-networks in the supplementary material and highlight the most overlapping regions and the shared node numbers ($N_{m}$) in Table \ref{table:real1}. Consistent with previous observations \citep{clinical1}, the abstinence related network anatomies are complex and occupy connections across different lobes. When further checking the involved functional systems, we conclude that the three sub-networks with their composed connections positively contribute to the NIE are  among corticolimbic (sub-network 1),  premotor (sub-network 13) and somatomotor (sub-network 14) systems; and the additional sub-networks with connections negatively contribute to the NIE are concentrated within the cerebellar (sub-network 4), frontoparietal (sub-network 7), prefrontal limbic (sub-network 10) and limbic (sub-network 12) systems. These findings are consistent with recent work in addictions. In particular, recent reviews have highlighted the important role of sensorimotor connectivity in the pathophysiology of addiction, suggesting that changes in this circuitry underly the transition from goal-directed to habitual drug use behavior \citep{gremel2017associative,yalachkov2010sensory}. Congruently, recent work in this same sample also identified sensorimotor, frontoparietal, cerebellar and subcortical connections as key predictors of opioid abstinence \citep{lichenstein2019dissociable}.

\section{Discussion}\label{sec:dis}
In this paper, we make the very first attempt to propose a mediation analysis under network-variate mediators. Motivated by an OUD study on how brain functional connectomes impact the treatment effect on abstinence, we develop a unified Bayesian mediation analysis named BNMM with latent network mediators that characterize  connectivity patterns across sub-networks as mediating effects. Without pre-specifying sub-network structures, we estimate modular allocations along joint modeling to facilitate a more supervised sub-network construction with enhanced mediating effects. Through a simultaneous feature selection based on sparse regressions, we are able to identify truly active mediating network components; and stabilize the model fitting with the conditional independence among latent network mediators. We show the superiority of our method in uncovering effect mechanisms by numerical studies.  

Our current method is based on a pre-determination of the number of latent blocks $Q$ in the SBMs based on the ICL criterion. Such a fixed block number is quite common in the general use of SBMs and has been frequently adopted when applying SBMs on brain functional connectome data \citep{zhang2019mixed,pavlovic2020multi} given that the potential number of blocks within a functional network is typically limited under a certain range in light of the brain functional architecture. To remove such a constraint and simultaneously derive block number from the data, we could replace the Dirichlet distribution of $\bfpi$ with a nonparametric Dirichlet process (DP) on the node-specific probability $\bfpi_v$. The discrete nature of the DP model will facilitate a natural grouping of the nodes, and in the posterior inference, we could resort to an approximate Gibbs sampler under the truncated stick-breaking process representation \citep{ishwaran2001gibbs,li2015spatial} with an efficient computation.

While our analysis is motivated by the mediating role of brain functional connectome, the modeling scheme is general to accommodate network-variate mediators and readily applicable to other application fields. For instance, it is common in genomics to consider gene pathways or networks in a graph format. Recently, mediation modeling with gene expression mediators has started to be introduced in eQTL analyses and other clinical studies \citep{shan2019identification}. A natural follow-up is to further investigate the mediation effects under genomic networks. Meanwhile, within the context of brain imaging, an additional natural extension would be to study mediation effects involving other types of brain networks including structural and multi-modal brain networks. We want to emphasize that the SBM representation for network mediators in our current setting is leveraged by the biological insight of the functional connectome. Under an alternative application, we could replace it with other low-rank parametric forms. For instance, we could resort to clique graph representation when modeling brain structural connectomes given that the anatomical neural supports are more concentrated.

Our current model focuses on a single exposure. With recent attempts on incorporating multivariate exposures \citep{wang2019bayesian} and the growing complexity of the collected data, extending BNMM to accommodate multiple or high dimensional exposure variables is a potential future direction to further generalize our method. For the outcome, we consider a continuous variable here and the regression link function could be replaced to handle other outcome types \citep{derkach2019high}. Some more demanding extensions, on which we are working, involve studying the network mediation for survival and longitudinal outcomes, which could offer additional perspectives on effect mechanisms. 

\section{Software}
\label{sec5}

The software for the proposed method in the form of R code is available at Github: \url{https://github.com/yizekaren/Bayesian_Network_Mediation_Model} to generate the simulated data and implement the model. 


\bibliographystyle{agsm}
\bibliography{refs,wileyNJD-AMA}

\begin{table}
 \caption{Simulation results for selecting truly active mediators and estimating effect components under different scenarios varying by the signal pattern and strength of noise. The Monte Carlo standard deviation for each evaluation metric is
included in the parentheses.}
 \begin{center}
\resizebox{\textwidth}{!}{
  \begin{tabular}{llrrrrr} 
  \toprule

 &  &Sensitivity & Specificity & Bias of NDE& Bias of NIE & Bias of TE\\

  \midrule
  \multicolumn{7}{c}{\textbf{\textit{Scenario 1}}}\\
  \hline
 \multirow{5}{*}{Low noise} & BNMM & \textbf{1.00} (0.02) & \textbf{1.00} (0.00) & \textbf{46.47} (0.49) & \textbf{-2.53} (0.49) & \textbf{-0.04} (0.08) \\
&   UMLM & 0.69 (0.20) & 0.99 (0.00) & 1836.53 (0.11) & -29.58 (6.38) & 65.32 (6.31) \\
 &  BAMALM & 0.76 (0.30) & \textbf{1.00} (0.00) & 336.20 (7.43) & 128.95 (34.93) & 139.49 (28.45) \\
&   UMBC & 0.63 (0.16) & 0.95 (0.01) &1840.47 (0.11) &5254.43 (508.15) & 5080.84 (508.08) \\
 &  BAMABC & 0.12 (0.17) & \textbf{1.00} (0.00) & 560.33 (5.21) & -76.40 (5.32) & -100.42 (5.21)\\

  \hline
 \multirow{5}{*}{High noise} &   BNMM & \textbf{0.92} (0.11) & \textbf{1.00}  (0.01) & \textbf{-74.73} (3.09) & \textbf{3.74} (3.16) & \textbf{-0.25} (0.41) \\
& UMLM  & 0.63 (0.20) & \textbf{1.00}  (0.00) & 1833.87 (0.40) & -33.45 (6.53) & 61.50 (6.52) \\
&   BAMALM  & 0.29 (0.35) & \textbf{1.00} (0.00) & 502.53 (5.69) & -33.89 (28.44) & -6.61 (25.20)\\
 & UMBC & 0.58 (0.17) & 0.96 (0.01) & 1837.73 (0.40) & 5013.57 (518.10) & 4852.10 (518.07) \\
 & BAMABC & 0.01 (0.07) & \textbf{1.00} (0.00) & 403.40 (4.59) & -80.12 (0.00) & -94.40 (4.59) \\

  \hline
  
  \hline
  \multicolumn{7}{c}{\textbf{\textit{Scenario 2}}}\\

  \hline
 \multirow{5}{*}{Low noise} &  BNMM & \textbf{0.98} (0.07) & 0.99 (0.01) & \textbf{36.40} (2.36) & \textbf{-24.91} (1.85) & \textbf{-21.43} (0.85)  \\
&   UMLM & 0.66 (0.22) & \textbf{1.00}  (0.01) & 1593.13 (0.12) & -32.08 (6.39) & 62.44 (6.31)\\
&  BAMALM & 0.61 (0.38) & \textbf{1.00} (0.01) & 217.53 (6.26) & 52.08 (27.99) & 61.71 (24.73)\\
&  UMBC & 0.59 (0.18) & 0.97 (0.01) & 1605.53 (0.09) & 2913.51 (331.07) & 2837.43 (333.04)\\
& BAMABC & 0.14 (0.12) & \textbf{1.00} (0.00) & 758.67 (4.32) & -87.50 (0.00) & -63.06 (4.32)\\

  \hline
 \multirow{5}{*}{High noise} & BNMM & \textbf{0.86} (0.17) & 0.98 (0.01) & 544.60 (4.28) & -48.25 (3.32) & \textbf{-13.77} (1.66) \\
&  UMLM & 0.61 (0.23) & \textbf{1.00}  (0.01) & 1592.47 (0.43) & \textbf{-35.39} (6.54) & 59.29 (6.49)\\
 & BAMALM & 0.11 (0.24) & \textbf{1.00} (0.00) & \textbf{245.53} (4.61) & -78.07 (13.99) & -59.252 (13.45)\\
&  UMBC & 0.55 (0.18) & 0.97 (0.01) & 1604.93 (0.43) & 2793.22 (345.32) & 2724.11 (345.36)\\
 & BAMABC & 0.10 (0.11) & \textbf{1.00} (0.00) & 535.13 (4.80) & -76.20 (0.00) & -96.06 (4.80)\\

\bottomrule
  
  \end{tabular}}
  \end{center}
  \label{table:simulation}
 \end{table}

\begin{table}
 \caption{Established sub-networks within the mediation analyses and their overlaps with canonical brain networks and macroscale regions. The sub-networks contribute to the truly active latent mediators are bolded.}\label{table:real1}
\resizebox{\textwidth}{!}{
\begin{tabular}{cccccc}
  \toprule
Sub-network & \# of nodes &  $N_{c}$  &         Canonical network                         &  $N_{m}$  &   Macroscale   regions                                \\
\midrule
\textbf{1}          & \textbf{21}   & \textbf{8}  & \textbf{Frontal-Parietal }               & \textbf{11} & \textbf{L-Prefrontal}                         \\
2          & 18   & 13 & Default Mode                     & 6  & L-Prefrontal/L-Limbic                \\
3          & 17   & 5  & Cingular-opercular/ Default mode & 4  & R-Temporal                           \\
\textbf{4}          & \textbf{12}   & \textbf{1}  & \textbf{Default Mode}                     & \textbf{6}  & \textbf{R-Cerebellum/L-Cerebellum}            \\
5          & 18   & 17 & Visual                           & 8  & L-Occipital                          \\
6          & 17   & 8  & Visual                           & 4  & L-Occipital                          \\
\textbf{7}          & \textbf{21}   & \textbf{10} & \textbf{Default Mode}                     & \textbf{7}  & \textbf{L-Prefrontal}                         \\
8          & 34   & 7  & Default Mode                     & 8  & L-Temporal                           \\
9          & 31   & 4  & Default Mode                     & 12 & R-Cerebellum                         \\
\textbf{10}         & \textbf{23}  & \textbf{9}  & \textbf{Subcortical}                      & \textbf{5}  & \textbf{R-Subcortical}                        \\
11         & 16   & 7  & Auditory                         & 3  & L-MotorStrip/L-Parietal              \\
\textbf{12}         & \textbf{20}   & \textbf{2}  & \textbf{Default Mode}                     & \textbf{4}  & \textbf{L-Limbic}                             \\
\textbf{13}         & \textbf{7}    & \textbf{6}  & \textbf{Somato-Motor}                     & \textbf{2}  & \textbf{L-MotorStrip/L-Parietal/R-MotorStrip} \\
\textbf{14}         & \textbf{13}   & \textbf{11} & \textbf{Somato-Motor}                     & \textbf{3}  & \textbf{L-Parietal/R-Parietal}               \\
\bottomrule
\end{tabular}}
\end{table}

 \newpage
 
\end{sloppypar}
\end{document}